\documentclass[multphys,vecphys]{svmult}

\usepackage{makeidx}     
\usepackage{graphicx}    
\usepackage{multicol}    

\makeindex             

\begin{document}

\title*{Expected Changes of SNe with Redshift due to Evolution of their Progenitors}
\titlerunning{Expected changes of SNe with redshift}
\author{Inma Dom\'\i nguez, \inst{1}
Peter H\"oflich, \inst{2}
Oscar Straniero, \inst{3}
Marco Limongi\inst{4}\and
Alessandro Chieffi\inst{5}}
\authorrunning{Dom\'\i nguez et al.} 
\institute{Universidad de Granada, Granada, Spain \texttt{inma@ugr.es} 
\and Universidad de Texas, Austin, USA
\and INAF-Osservatorio Astronomico di Collurania, Teramo, Italy
\and INAF-Osservatorio Astronomico di Roma, Monteporzio, Italy
\and Istituto di Astrofisica Spaziale, CNR, Roma, Italy}
%
%
\maketitle

\abstract

We have analyzed the influence of the stellar populations, from which
 SN progenitors come from, on the observational outcome, including the metal
 free Pop. III. We use our
 models to study the evolution of the progenitor, the subsequent explosion and
 the light curves. 
 
 For Type Ia, the variation of the main sequence mass of the progenitor of the
 exploding WD, produces an offset in the maximum-decline relation of 0.2 mag. 
  This effect is critical for 
 the use of high redshift Type Ia SNe as cosmological standard candles. 
  In contrast, the metallicity does not change the above relation (at maximum, $\Delta M_V \le$0.06 mag). 
 
 For Type II, we find a
 dependence of the light curve properties with both, main sequence mass and metallicity of the progenitor, 
 and we identify a rather homogeneous subclass, {\it Extreme} II-P, 
 that may be used as a quasi-standard candle. Note that, although not as good as Type Ia for distance 
  determinations, Type II are expected to occur since the first stars were formed. 

\section{Introduction}

 Due to their high brightness, SNe are among the best candidates to measure distances on cosmological scales;   
in particular, Type Ia SNe (SNe~Ia) are the favorites as they are brighter by about 1 to 2 magnitudes than other class of SNe 
  and show very
homogeneous properties.
  Recently, SNe~Ia observed at
 high redshifts, z, (see \cite{S98} and \cite{P99})  
 have provided results that are
 consistent with a Universe composed mainly of  {\it dark energy} or, equivalent, with a positive
 cosmological constant, $\Omega_\Lambda $  $\sim 0.7$.
Note that this conclusion relies
 on the application of an empirical relation between light curve (LC) shape and maximum luminosity 
 which is obtained for the well observed nearby SNe Ia occurring in  galaxies
 at known distance ( \cite{P99} and \cite{R96}). 
 
 To understand the nature of this  
 {\it dark energy}, we need to observe standard-candles at high z.
 The quantity and quality of the data 
 is increasing due to systematic searches, space telescopes and 8m-class ground telescopes. 
However, SNe or other standard-candles are calibrated locally, and their application to very high z  
 may be limited by evolutionary effects.  
In fact, there is observational evidence indicating a correlation of SN properties
with those of their host galaxies (\cite{B96}, \cite{Ca97}, \cite{Ha00}, \cite{IHP00} and \cite{WHW97}).

 SNe~Ia are thought to
 be the  thermonuclear explosion of CO white-dwarfs \cite{HF60}
  triggered by  mass accretion from
 a companion.  The proposed scenarios are merging of WDs by gravitational
 radiation  (double degenarate) \cite{We84} and \cite{IT84}, and
 accretion of H and/or He from a non-degenerate companion (single degenerate) \cite{WI73}. 
 Moreover, they occur in elliptical galaxies indicating that some of their 
  progenitors, at least, have evolutionary time-scales comparable to the Hubble time and, thus,  at high z SNe~Ia should be rare.

The other class of SNe, core collapse SNe, are thought to be the explosion 
 of massive stars (greater than $\sim 10 M_\odot$) caused by the collapse of its central parts into a neutron star or a black hole. Their evolutionary time scales are short compared to the age of the universe even at
high z (\cite{T82}, \cite{MB76} and \cite{WW86}).
Core collapse SNe show a wide range of brightness, up to 6 magnitudes, and
 properties of their light curves (\cite{Fi00}, \cite{Pa94} and \cite{YB89}) 
 which prevents their use as standard-candles. However,
 our knowledge of the event is improving and it may be possible to derive
 the absolute magnitude in a similar way as Type Ia if appropriate empirical
 correlations can be identified (see Hamuy, this volume and \cite{HP02}).
 These objects will occur soon after the initial star formation period and, therefore, can be used
to probe the structure of the universe at high z.
 E.g. at $z \sim 5$, galaxies are expected to be small and dim and
core collapse supernovae may be the brightest objects in the Universe \cite{MR97}.

It is expected that, going back in time, the stellar population from which SN   
 progenitors come out, would be composed of more low metallicity and rapid evolving, more massive, 
 stars.
In this work we focus, mainly, on the exploration of the sensitivity 
 of the light curve characteristics on the underlying progenitor properties: initial mass and metallicity (Z). 
 
\section{Numerical Models} 

\subsection{Stellar Evolution}

The evolution of selected models in the mass range from 1.5 to 25 M$_\odot$
 and Z between 0 (Pop.III) and 0.02 (solar) have been computed from the pre-main sequence
 to the TP-AGB phase in the case of the low and intermediate mass stars and to the onset
 of core collapse for the massive stars. We have not included rotation neither mass loss.  

All models have been computed by means
of the evolutionary code FRANEC (rel 4.2) and are  
 extensively described in \cite{DHS01}, \cite{LSC00}   
 and \cite{IIP03}.
The details of the FRANEC has been
presented in \cite{CS89}, \cite{SCL97} and \cite{CLS98}.

\subsection{Explosions and Light Curves}

Based on the previous models, the explosion, detailed
 post-processing and light curves  are computed by means of
 a 1D radiation-hydrodynamic code (see \cite{HK96}, \cite{DHS01} and \cite{IIP03}).
All parameters are fixed in order to analyze the sensitivity of 
 the observed properties to the initial mass and metallicity of the progenitors. 

For Type Ia we have considered delayed detonation explosions because models based on this 
 explosion mechanism reproduce monochromatic light curves and spectra reasonable well, 
 including the maximum-LC shape relation (see for example 
 \cite{Ho95}, \cite{HKW95}, \cite{Le00}, \cite{Nu97} and \cite{Wh98}). The description of the velocity of the deflagration front is
    based on 3D
simulations \cite{DH00}, \cite{Kh95} and \cite{Ni95}; model parameters, transition density and ignition density,
 are fixed and have been chosen to reproduce a typical SN Ia.   

For Type II, the explosion is triggered artificially by depositing energy above the mass cut, 
 fullfilling the requirement that the final kinetic energy and the $^{56}$Ni mass is the same in all models.   

\section{Results}

\subsection{Type Ia}

We have connected the initial mass and metallicity of the WD progenitor to the light curves and spectral 
 properties of SNe Ia. All the potential progenitors of the WD have been studied. The key parameter is the 
  integrated C/O ratio, the fuel, within the Chadrasekhar mass 
  WD. In general, changes are small 
  because the nuclear energy released by a complete burning (into iron group ashes)
 of a pure Carbon WD is about 10 $\%$ larger than that released by a pure Oxygen WD with
 the same mass.

We find that the initial mass of the progenitor modifies the average C/O in the WD up to a 22 $\%$ and, 
 as a consequence, the $^{56}$Ni mass produced by the explosion and the
 kinetic energy (for details see \cite{DHS01} and also \cite{Ho98}). Progenitors with greater masses 
  (smaller C/O rate within the WD and consequently less amount of $^{56}$Ni) produce less luminous and slightly 
 faster decline LCs. In particular, the initial mass alters the maximum-LC shape relation, which may be 
  offset by up to 0.2 mag. Notice that for a detailed analysis of the nature of the {\it dark} energy an 
   accuracy of 0.05 mag is required \cite{Al00}.
   
 In addition, for these massive progenitors, kinetic energies and expansion velocities are also smaller, 
 down to 2000 km/s. This correlation between LC shape and expansion velocity may be used to reduce 
 the scatter in the empirical relation from which the maximum luminosity is obtained.

The dependence of the total C/O on the initial mass is mainly due to the
different extension of the convective core during the central He burning
phase. In fact, the internal structure of an exploding WD may be
schematically divided in
two distinct regions: an internal one, where C is significantly depleted
(about 25 $\%$ C and 75 $\%$ O) and an external one, where C/O is about 1. The
internal region is built in during the core He-burning phase, while the
external one is left by the shell burning  (AGB or accretion phase).
Since the extension of the central region (C depleted) coincides with
the convective core  of the He-burning progenitor, and since more
massive progenitors have a larger convective cores, the larger is the
progenitor mass, the smaller is the final (pre-explosive) C/O.

In contrast, the dependence of the size of this inner C-depleted region with metallicity is much 
 weaker and not monotonic.  In fact, the metallicity of the progenitor does not influence the
average C/O ratio within the WD; changes are smaller than 9 $\%$ and, as a consequence,  
 the amount of $^{56}$Ni produced in the explosion and the kinetic energies
are rather similar (but see \cite{Ho00}).
For this reason, metallicity does not influence the maximum luminosity ($\Delta M_V \le$0.06 mag), 
nor the LC shape. 

Finally, we want to stress, that the final integrated C/O ratio within the pre-explosive Chandrasekhar 
 mass WD depends on the scenario; mergers (double degenerate) are expected to have a smaller C/O ratio 
  compared to single degenerates. The reason is that, in the case of merging, two central carbon depleted region 
   corresponding to the two WDs, are added.    

\subsection{Type II}

First, we want to mention the limitation of our theoretical study; we have explored a very limited parameter 
space (see Hamuy, this volume, and \cite{Ha03}) 
  and to compare with 
 observations, different $^{56}$Ni masses and kinetic energies should be considered and 
  mass loss included during the evolution of the 
  progenitor. 

We find (see \cite{H00} and \cite{IIP03} for details) that all the Pop. III models end up as Blue Super Giants (BSG) while all the solar
metallicity ones end up as Red Super Giants (RSG). At intermediate Z, 
the more massive stars end up as BSG while the less massive ones end up as RSG.  The limiting
mass depends on Z. 
 In general, the stellar
 radius depends on the opacity of the envelope; the lower the metallicity, the
 lower the opacity and, as a consequence, the structure is more compact.

If the progenitor is a RSG, the brightness during the plateau phase,
 which lasts more than 50 days,  is nearly
 constant, $M_V \sim$ -17.5, and quite insensitive to changes of the initial mass  
 ($\Delta M_V \le$0.07 mag). Increasing the kinetic energy by a factor of 2, results in a 
  slightly brighter plateau, $\Delta M_V \le $0.3 mag. This makes this
 {\it sub-class}, which we call {\it Extreme} II-P, a quasi-standard candle. 
 
 The similarity of the LCs is caused by the
  similarity  of the density structures
of the RSG envelopes, which show flat density gradients at the the photosphere, which during 
 the plateau phase is 
 located at the H-recombination front. A self-regulating mechanism is at work: an increase in
the energy release causes a heating of the photosphere and a reduction
of the energy production rate and viceversa.
 Because the density slope of the photosphere is flat and slowly changing,
the luminosity profile of the light curves is rather flat.

Moreover, the unique light curves of {\it Extreme} II-P 
 allow photometric identification and so, permit their observation from ground 8m-class
 telescopes up to a redshift of z$\sim$ 3. Space telescopes, like SIRTF, would push this limit  even further and the NGST 
 is expected to detect SNII up to z$\sim$ 10.

BSG progenitors produce sub-luminous
 events, 1.5 magnitudes fainter than  SNII which come from RSGs. The light curve presents, 
 instead of the plateau, a long-lasting phase of
 increasing brightness. In this case the self-regulating mechanism does not work due 
 to the steep density gradients.  
   
 Finally, these studies are very preliminary, key problems, concerning
 the identification of the progenitors (Type Ia) and the explosion mechanisms
 (Type Ia and Type II), are still to be solved, and, besides, we do not know how the
 stellar populations evolve with redshift.


\begin{thebibliography}{}
%
%
%
\bibitem{Al00}
 Albrecht, A., \& Weller, J.: AAS \textbf{197}, 6106 (2000)
\bibitem{B96}
 Branch, D., Romanishing, W. \& Baron, E.: ApJ \textbf{465}, 73 (1996)
\bibitem{Ca97}
 Cappellaro, E. et al.: A\&A \textbf{322}, 421 (1997)
\bibitem{CS89} Chieffi, A., and Straniero, O.: ApJS \textbf{71}, 48 (1989)
\bibitem{CLS98} Chieffi, A., Limongi, M., and Straniero, O.: ApJ \textbf{502}, 737 (1998)
\bibitem{CL02} Chieffi, A., and Limongi, M.: ApJ \textbf{577}, 281 (2002)
\bibitem{IIP03} Chieffi, A., Dom\'\i nguez, I., H\"{o}flich, P., Limongi, M., Straniero, O.:
  MNRAS, in press (2003)
\bibitem{DH00}
 Dom\'\i nguez, I.,  H\"{o}flich, P.: ApJ \textbf{528}, 854 (2000)
\bibitem{DHS01}
 Dom\'\i nguez, I.,  H\"{o}flich, P., Straniero, O.: ApJ \textbf{557}, 279 (2001)
\bibitem{Fi00} Filippenko A.V.: {\it Cosmic Explosions}. Ed. by S. Holt
 \& W.W. Zhang, New York, American Institute of Physics (2000)
\bibitem{Ha00}
 Hamuy, M. et al: AJ \textbf{120}, 1479 (2000)
\bibitem{HP02} Hamuy, M., and Pinto,  P.A.:  ApJ \textbf{566}, L63 (2002)
\bibitem{Ha03} Hamuy, M.: ApJ, \textbf{582}, 905 (2003)
\bibitem{Ho95} H\"oflich, P.: ApJ \textbf{443}, 89 (1995)
\bibitem{HKW95} H\"oflich, P., Khokhlov, A and Wheeler J.C.: ApJ \textbf{444}, 831 (1995)
\bibitem{HK96} H\"oflich P., Khokhlov A.: ApJ \textbf{457}, 500 (1996)
\bibitem{Ho98} H\"oflich, P., Wheeler J.C., Thielemann F.K.: ApJ \textbf{495}, 617 (1998)
\bibitem{Ho00} H\"oflich, P., Nomoto, K., Umeda, H.,  and Wheeler, J.C.: ApJ  \textbf{528}, 590 (2000)
\bibitem{H00} H\"oflich P., Straniero O., Limongi M., Dom\'\i nguez I., Chieffi A.: {\it  7th Texas-Mexican Conference}.  Ed by W. Lee \& S. Torres-Peimbert, Rev. Mex. A\&A, Vol. 10, p. 157 (2000)
\bibitem{HF60} Hoyle P., Fowler W.A.: ApJ, \textbf{132}, 565 (1960)
\bibitem{IT84} Iben I.J, Tutukov A.V.: ApJS  \textbf{54}, 335 (1984)
\bibitem{IHP00} Ivanov, V.D., Hamuy, M., Pinto, P.A.: ApJ \textbf{542}, 588 (2000)
\bibitem{Kh95} Khokhlov A.: ApJ \textbf{457}, 695 (1995)
\bibitem{Le00} Lentz, E.J., Baron, E., Branch, D., Hauschildt, P., Nugent, P.E.:  ApJ \textbf{530}, 966 (2000) 
\bibitem{LCS01} Limongi M., Chieffi A., Straniero O.: ApJS \textbf{129}, 625L (2001)
\bibitem{LSC00} Limongi, M., Straniero, O., Chieffi, A.: ApJS \textbf{129}, 625 (2000)
\bibitem{MB76} Maza J., van den Bergh, S.: ApJ~ \textbf{204}, 519 (1976)
\bibitem{MR97} Miralda-Escud\'e, J., Rees, M.J.: ApJ \textbf{478}, 57 (1997)
\bibitem{Ni95} Niemeyer, J.C., Hillebrandt, W.:  ApJ \textbf{452}, 779 (1995)
\bibitem{Nu97} Nugent, P.E., Baron, E., Hauschildt, P.,  Branch, D.:  ApJ \textbf{485}, 812 (1997)
\bibitem{Pa94} Patat F., Barbon R., Capellaro E., Turatto M.: A\&A ~ \textbf{282}, 731 (1994)
\bibitem{P99} Perlmutter, S. et al: ApJ  \textbf{517}, 565 (1999)
\bibitem{Ph87} Phillips  M.M. et al.: { PASP} \textbf{ 90}, 592 (1987)
\bibitem{Ph99} Phillips  M.M., Lira P., Suntzeff N.B., Schommer R.A., Hamuy M., Maza J.: AJ \textbf{118}, 1766 (1999) 
\bibitem{R96} Riess A.G., Press W.H., Kirshner R.P.: ApJ \textbf{473}, 588 (1996)
\bibitem{S98} Schmidt, B. P. et al: ApJ  \textbf{507}, 46 (1998)
\bibitem{SCL97} Straniero, O., Chieffi, A., and Limongi, M.: ApJ \textbf{490}, 425 (1997)
\bibitem{T82} Tammann G.: {\it NATO-ASI on Supernovae: A Survey of Current
 Research.} Ed. by M.J. Rees and R.J. Stoneham, p. 371, Reidel, Dordrecht (1982)
\bibitem{WHW97} Wang, L., H\"oflich P., Wheeler, J.C.: ApJ \textbf{487},  L29 (1997)
\bibitem{We84}  Webbink  R.F.: {ApJ}, \textbf{277}, {355} (1984)
\bibitem{WI73} Whelan J., Iben I.Jr.: {ApJ} \textbf{186}, 1007 (1973)
\bibitem{Wh98}  Wheeler J.C., H\"oflich, P., Harkness, R., Spyromilio, J.: ApJ \textbf{496}, 908 (1998)
\bibitem{WW86} Woosley S.E., Weaver T.A.: ARA\&A ~ \textbf{24}, 205 (1986)
\bibitem{YB89} Young T.R., Branch D.: ApJ~ \textbf{342}, L79 (1989)

\end{thebibliography}
%


\printindex
\end{document}